\documentclass[prl,twocolumn,showkeys,preprintnumbers,floats,floatfix]{revtex4}

\usepackage{graphicx}
\usepackage{color}
\usepackage{soul}
\usepackage{amsmath}

\begin{document}

\title{Electrostatically defined few-electron double quantum dot in silicon}

\author{W.H. Lim}
\affiliation{Australian Research Council Centre of Excellence for Quantum Computer Technology, The University of New South Wales, Sydney 2052, Australia}
\author{H. Huebl}
\affiliation{Australian Research Council Centre of Excellence for Quantum Computer Technology, The University of New South Wales, Sydney 2052, Australia}
\author{L.H. Willems van Beveren}
\affiliation{Australian Research Council Centre of Excellence for Quantum Computer Technology, The University of New South Wales, Sydney 2052, Australia}
\author{S. Rubanov}
\affiliation{Australian Research Council Centre of Excellence for
Quantum Computer Technology, The University of Melbourne, Victoria
3010, Australia}
\author{P.G. Spizzirri}
\affiliation{Australian Research Council Centre of Excellence for Quantum Computer Technology, The University of Melbourne, Victoria 3010, Australia}
\author{S.J. Angus}
\affiliation{Australian Research Council Centre of Excellence for Quantum Computer Technology, The University of Melbourne, Victoria 3010, Australia}
\author{R.G. Clark}
\affiliation{Australian Research Council Centre of Excellence for Quantum Computer Technology, The University of New South Wales, Sydney 2052, Australia}
\author{A.S. Dzurak}
\affiliation{Australian Research Council Centre of Excellence for Quantum Computer Technology, The University of New South Wales, Sydney 2052, Australia}

\date{\today}

\begin{abstract}
A few-electron double quantum dot was fabricated using
metal-oxide-semiconductor(MOS)-compatible technology and
low-temperature transport measurements were performed to study the
energy spectrum of the device. The double dot structure is
electrically tunable, enabling the inter-dot coupling to be adjusted
over a wide range, as observed in the charge stability diagram.
Resonant single-electron tunneling through ground and excited states
of the double dot was clearly observed in bias spectroscopy
measurements.
\end{abstract}



\maketitle

Electrostatically defined single and double quantum dot (DQD)
systems in GaAs/AlGaAs heterostructures~\cite{Chan2004,Wiel2002} are
the current benchmark for the implementation of DiVincenzo's
criteria using semiconductor
qubits~\cite{Loss1998,Petta2005,Koppens2006}. Although the nuclear
spins inherently present in GaAs provide a fast decoherence
mechanism, this drawback has been partly overcome
recently~\cite{Reilly2008}. Silicon has a natural advantage in this
respect since the only stable isotope with a nuclear spin is
$^{29}$Si. The 4.7$\%$ abundance of this isotope in
$^{\textrm{nat}}$Si can be reduced by isotopic purification,
resulting in nearly nuclear-spin-free crystals. This should, in
principle, increase the coherence time of electron-spin qubits in
silicon~\cite{Tahan2002,Tyryshkin2006}. Initial demonstrations of
Si-based DQD systems for spin qubits~\cite{Lee2006,Shin2007} have
stimulated a number of recent studies of DQDs in both multi-gated
silicon-on-insulator (SOI)~\cite{Liu2008,HongwuLiu2008} and
Si/SiGe~\cite{Shaji2008} structures.

In this letter, we report the fabrication of a few-electron DQD and
its electrical measurement at milli-kelvin temperatures. The double
dot is based upon a recently developed double-gated silicon quantum
dot~\cite{Angus2007}, which was also shown to operate effectively as
a radio-frequency single electron transistor~\cite{Angus2008}. Our
approach provides a simple method of producing multi-gated silicon
quantum dots without the need for complementary-MOS (CMOS) process
technologies, such as polysilicon deposition and etching. The
morphology of the double dot device is investigated using
cross-sectional transmission electron microscopy (XTEM) analysis.
Transport spectroscopy demonstrates the ability to tune the double
dot from the weakly-coupled to strongly-coupled regime. In the
weakly-coupled regime, extracted capacitances of the system show
good quantitative agreement with simple modelling using
FastCap~\cite{Nabors1991}.

\begin{figure}[t]
\includegraphics[width=8.5cm]{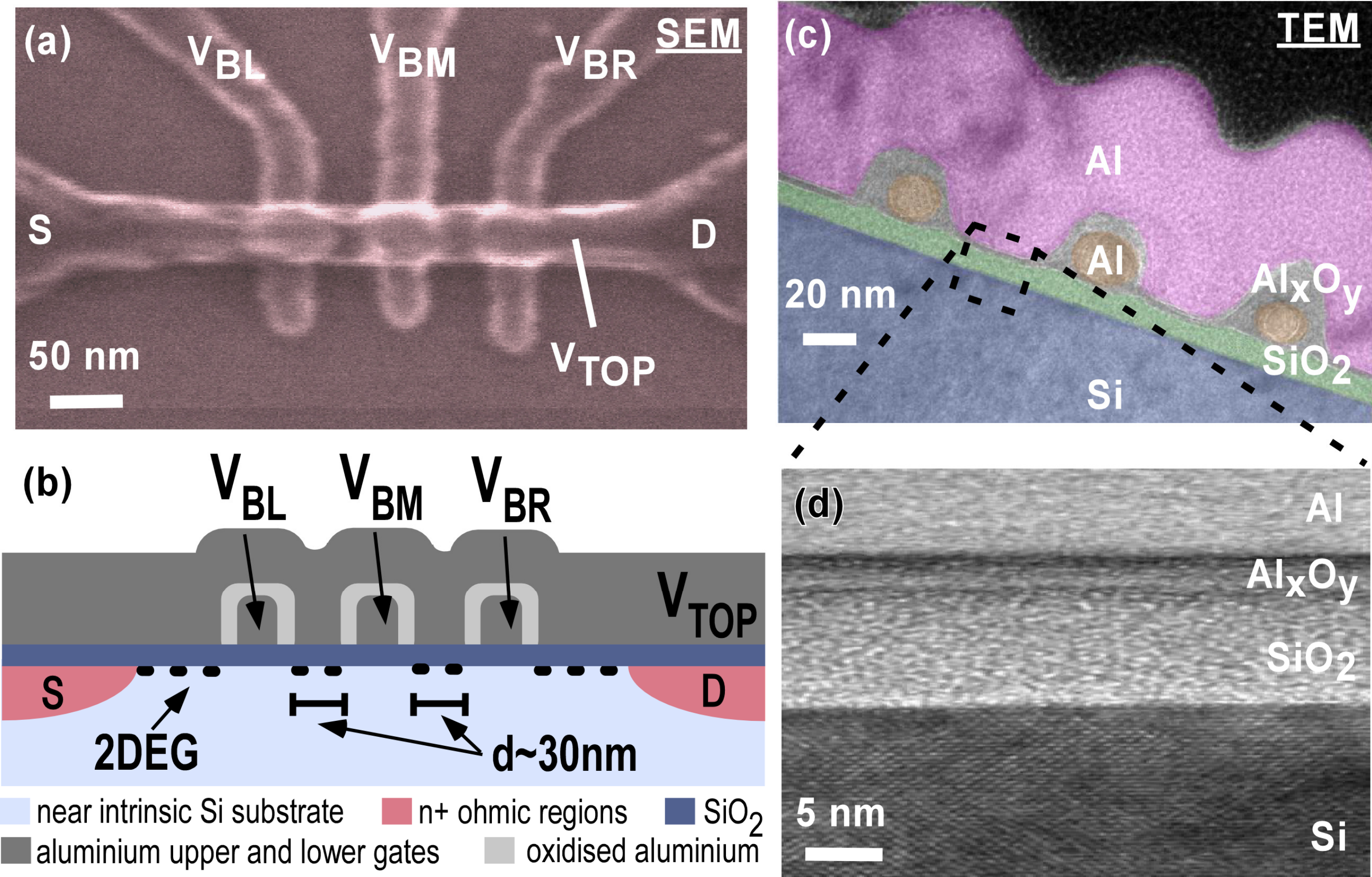}
\caption{(a) SEM image of the Si MOS DQD. The three barrier gates and the top gate have widths $\sim$30 nm and $\sim$50 nm respectively. The Al barrier gates were plasma-oxidized to isolate them from the top gate. (b) Schematic cross-section of the device (not to scale). Source and drain n$^+$ contacts (red) were formed by phosphorus diffusion into the Si substrate (light blue). The top gate induces a 2DEG and the barrier gates create three potential barriers, forming two dots. The size of the dots is estimated to be $30\times50$ nm$^2$. (c) Color-enhanced XTEM image of a similar device. (d) Enlarged XTEM image, showing sharp interfaces between the Si substrate, SiO$_2$ gate oxide, Al$_x$O$_y$ and the Al top gate.}
\label{fig1}
\end{figure}

The devices investigated in this work were fabricated on
near-intrinsic silicon wafers ($\rho$ $>$ 10 k$\Omega\cdot$cm at
300~K). After definition of n$^+$ ohmic contacts by phosphorus
diffusion through a masked sacrificial thermal oxide, a 200~nm field
oxide was grown. In the active device region
(30$\times$30~$\mu$m$^2$), the field oxide was etched locally and
replaced by an 8 nm-thick high-quality SiO$_2$ gate oxide, grown in
an ultra-dry oxidation furnace at 800 $^\circ$C in O$_2$ and
dichloroethylene. Three Al barrier gates were then patterned by
electron beam lithography (EBL), thermal evaporation and lift-off.
The barrier gates were next passivated by plasma
oxidation~\cite{Angus2007,Heij2001}, resulting in an
electrically-insulating Al$_x$O$_y$ layer surrounding the barrier
gates. The Al top gate was defined in a second EBL step aligned to
the lower gates with an accuracy of $\sim$20~nm. Finally, the
devices were annealed at 400~$^\circ$C for 15 mins in forming gas
(95$\%$ N$_2$/5$\%$ H$_2$) to reduce the Si/SiO$_2$ interface trap
density($D_{it}$). Deep-level transient spectroscopy of
similarly-processed structures revealed $D_{it}$ of order
5$\times$10$^{10}$ cm$^{-2}$eV$^{-1}$ near the conduction band
edge~\cite{McCallum2008}.

Figures 1(a,b) show a scanning electron microscope (SEM) image and a
schematic cross-section of a double dot device. The top gate, which
extends over the source and drain n$^+$ contacts (not shown) and
also the three barrier gates are used to form a two-dimensional
electron gas (2DEG) accumulation layer under the thin SiO$_2$ layer.
The barrier gates are used to locally deplete the 2DEG, forming
three tunnel barriers that define two dots in series. The dots are
geometrically defined by the distance between adjacent barrier gates
($\sim$30 nm), and by the top gate width ($\sim$50 nm). The outer
barrier gates and top gate are used to control the electron
occupancies electrostatically and the middle barrier gate is used to
control the inter-dot coupling.

Figure 1(c) shows an XTEM image along the top gate (i.e. perpendicular
to the barrier gates). Apart from an increased (200~nm) top-gate
width in order to aid XTEM sample preparation, this device is
nominally identical to the device used in electrical measurements.
The XTEM image confirms the target 5~nm Al$_x$O$_y$ layer thickness
from the plasma oxidation process used (100~mTorr, 50~W incident RF
O$_2$ plasma, 150~$^\circ$C for 3~mins). Interestingly, at the
interface between the top gate and the SiO$_2$, we find an
additional Al$_x$O$_y$ layer ($\sim$2~nm thick, see Fig. 1(d)) which
could be due to the oxidation of the Al top gate via chemical
interaction with the SiO$_2$ layer below. We note that the Al
barrier gates, initially evaporated to a thickness of 30 nm, show an
Al core of only $\sim$20~nm in diameter after plasma oxidation,
consistent with the formation of a $\sim$5 nm Al$_x$O$_y$ insulator.
This Al$_x$O$_y$ thickness is sufficient to allow differential
biases of up to 4 V between the upper and lower gates with
negligible leakage.

Electrical (dc) transport measurements were performed in an Oxford
Instruments Kelvinox K100 dilution refrigerator at a base
temperature of $\sim$50 mK. A source-drain excitation voltage
$V_{sd}$=50~$\mu$V at a modulation frequency of 13 Hz was used to
monitor the differential conductance $dI/dV_{sd}$. The source-drain
dc current $I_{SD}$ was measured with a room-temperature current
preamplifier. Initially, the left (right) dot was characterized
independently by setting the right (left) barrier-gate voltage
$V_{BR}$($V_{BL}$) equal to the top gate voltage $V_{TOP}$. The
middle barrier gate voltage  $V_{BM}$ was fixed at $V_{BM}$=0.818~V.
Under these conditions Coulomb diamonds were recorded and the
charging energy of the left (right) dot, was determined to be
$E_C\sim$5~meV ($\sim$2.5~meV) at $V_{TOP}$=1.6~V. Therefore, the
total capacitance of the left (right) dot was
$C_{\Sigma,left(right)}$=$e^2/E_C\sim$30~aF ($\sim$60~aF) at
$V_{BL}$=0.76~V ($V_{BR}$=0.76~V). To compare these experimentally
obtained results with modelled parameters, we used FastCap which
calculates the capacitances based on a finite element approach.
Using the lithographic device dimensions as inputs, we obtained a
total capacitance $C_{\Sigma}\sim$30 aF for both dots, in good
agreement with the experimental value for the left dot but at
variance with that of the right dot by a factor of two. Such
variations in capacitance from dot to dot could result from physical
asymmetries in real devices, as evidenced by the XTEM image in Fig.
1(c), or from the presence of fixed charge in the gate oxide or at
interfaces which can modify the effective gate potentials.

\begin{figure}[t]
\includegraphics[width=8.5cm]{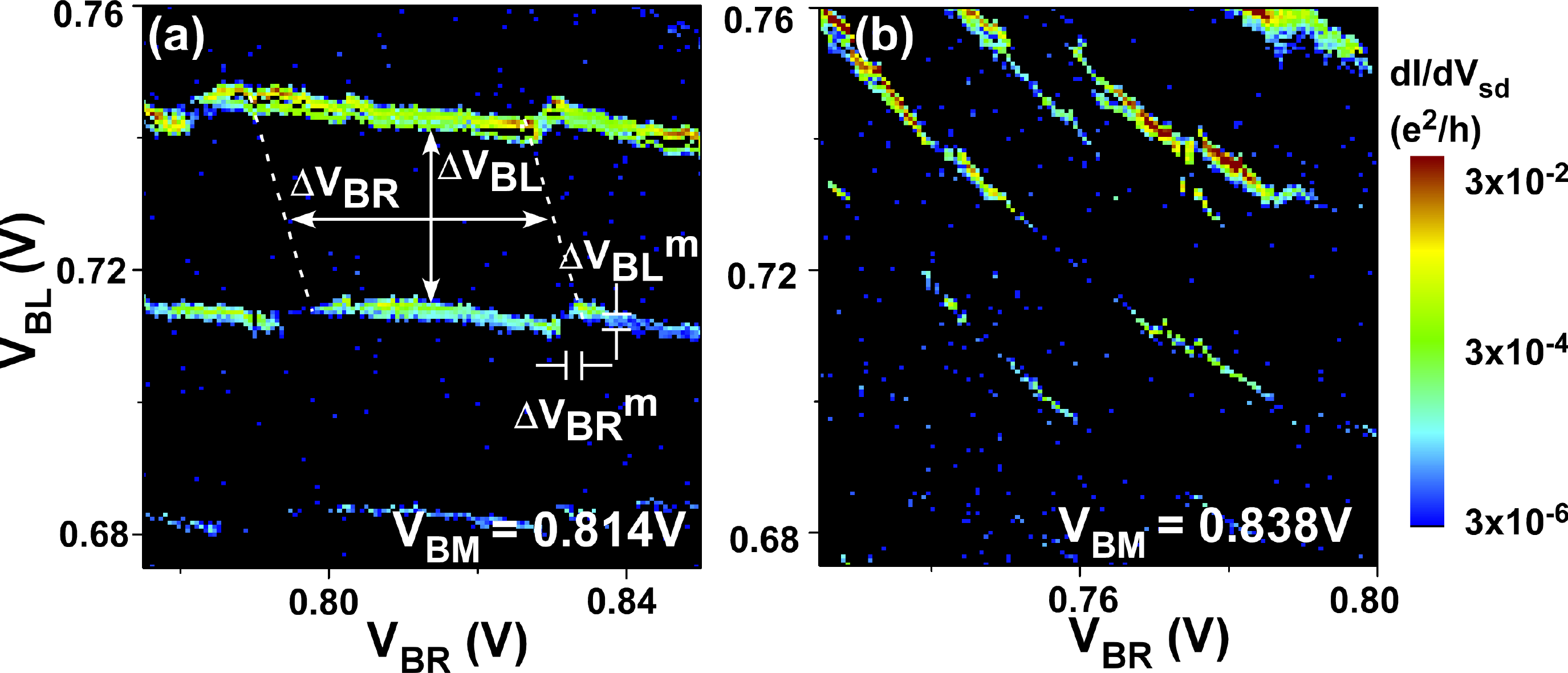}
\caption{Differential conductance $dI/dV_{sd}$ as a function of the barrier gate voltages, $V_{BL}$ and $V_{BR}$, for $V_{TOP}$=1.6 V and zero source-drain bias. Tuning the middle gate voltage in the range $V_{BM}$=0.814 V $-$ 0.830 V, we observe a transition from two almost isolated dots (a) to the formation of a single large dot (b).}
\label{fig2}
\end{figure}

We estimate the electron occupancy of a single dot using two
methods. The first method uses Hall measurements of a similar MOSFET
device from which the electron density is determined to be
$n=3.5\times(V_{TOP}- V_{TH}$)10$^{12}$
cm$^{-2}$~\cite{McCamey2007}, where $V_{TH}$ is the threshold
voltage. When operated as a simple MOSFET, our device showed
$V_{TH}\sim$1.25~V. Hence, at $V_{TOP}$=1.6~V we estimate the 2DEG
density of our device to be $n\sim1.2\times10^{12}$~cm$^{-2}$,
resulting in a dot occupancy of N$\sim$20 electrons for a
30$\times$50~nm dot size. As an alternative method, we estimate
electron occupancy by counting Coulomb oscillations from $V_{TH}$,
assuming no free electrons in the dots below threshold
voltage~\cite{Angus2007}. This method derives a dot occupancy of
N$\sim$15, in reasonable agreement with the previous method. Both
approaches indicate that the device operates in the few electron
regime.

Figure 2 shows the differential conductance $dI/dV_{sd}$ of the DQD
as a function of the barrier-gate voltages, $V_{BL}$ and $V_{BR}$,
for a fixed top-gate voltage $V_{TOP}$=1.6~V and source drain
voltage $V_{SD}$=0~V for two different middle barrier-gate voltages
$V_{BM}$. In Fig. 2(a), the relatively low middle barrier-gate
voltage $V_{BM}$=0.814~V and therefore high central barrier
separates the two dots, resulting in the characteristic
honeycomb-shaped charge stability diagram. By calculating the
voltage ratios
${\Delta}V_{BR}^m$/${\Delta}V_{BR}$ (${\Delta}V_{BL}^m$/${\Delta}V_{BL}$),
we can estimate the ratios of the mutual capacitance to the total
dot capacitance $C_m$/$C_{\Sigma,left(right)}\sim$0.10(0.07),
indicating that the double dot is in the weak coupling
regime~\cite{Wiel2002}. There, we observe the characteristic triple
points resulting from the alignment of the electrochemical
potentials of the dots and the leads. In addition, current is
observed along the sides of the hexagons, which can occur when the
dots are strongly coupled to the leads and second-order co-tunneling
processes occur~\cite{Francesni2001}. Increasing the middle barrier
gate voltage to $V_{BM}$=0.838 V, the mutual capacitance increases
and dominates the system ($C_m$/$C_{\Sigma,left(right)}\sim$1). This situation occurs when the
middle barrier is reduced and a single (merged) large dot is formed,
resulting in diagonal parallel Coulomb lines, as observed in Fig.
2(b).

Figure 3(a) shows transport data through the DQD in the weak
coupling regime $V_{BM}$=0.802 V with $V_{TOP}$=1.4~V and
$V_{SD}$=$-$1.0~mV. For $|V_{SD}|>$0 the triple points evolve into
so-called bias triangles, reflecting the occurrence of transport
within the bias windows~\cite{Wiel2002}. In a double dot system, two
types of coupling can be distinguished: capacitive coupling; and
tunnel coupling. While capacitive coupling is a purely classical
effect, tunnel coupling arises from the overlap of electron wave
functions, classified by the fractional splitting ratio
$F=2{\Delta}V_s/V_p$, where ${\Delta}V_s$ is the splitting between
the paired triangles and ${\Delta}V_p$ is the diagonal separation
between triangle pairs in Fig.~3(a)~\cite{Waugh1995,Mason2004}.
Here, we find $F\sim$0.2, indicating that the interaction between
the two dots is dominated by capacitive coupling. The device may
therefore be modelled using a capacitive approach.

\begin{figure}[t]
\includegraphics[width=9.0cm]{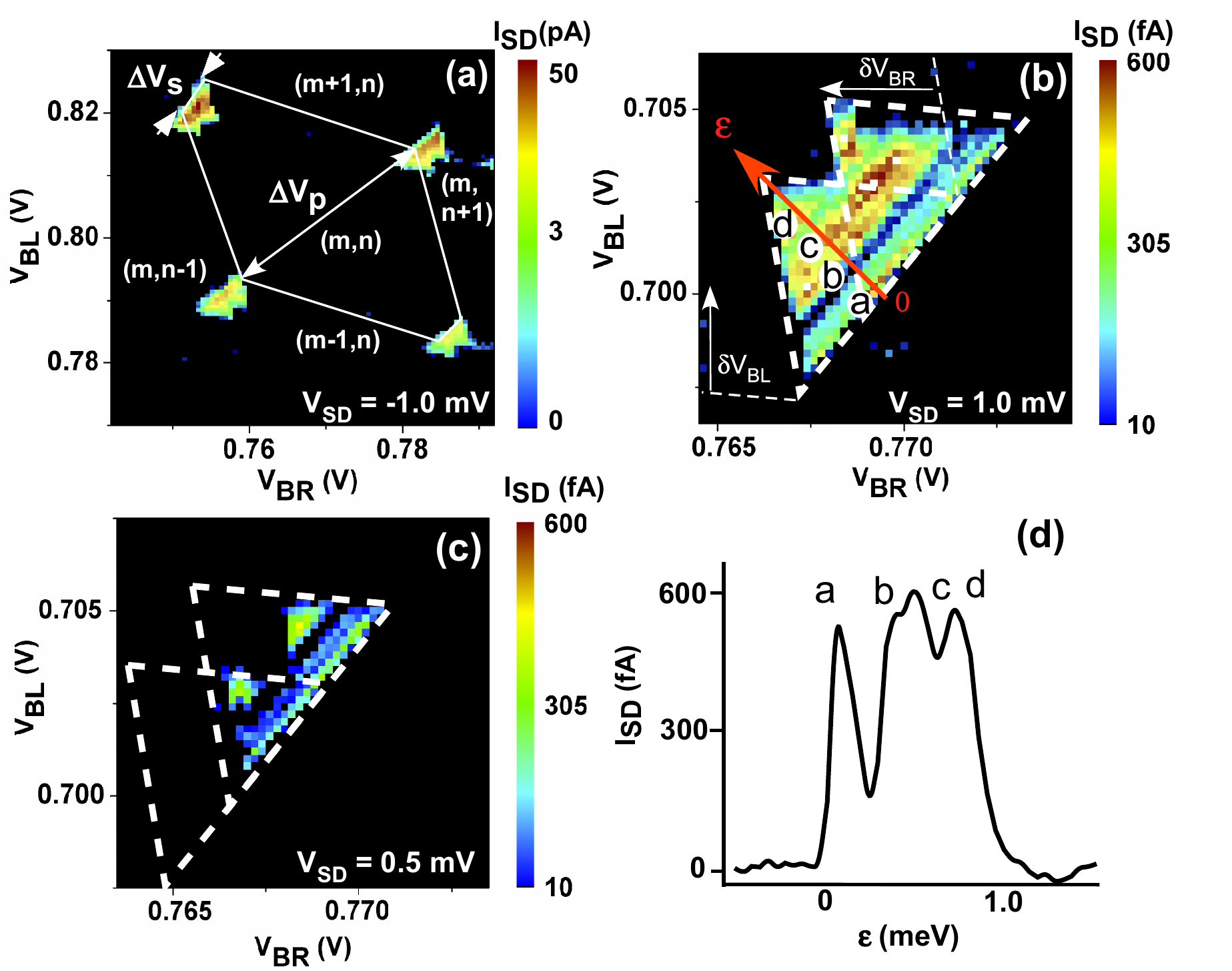}
\caption{Bias spectroscopy of a weakly coupled double dot with $V_{BM}$=0.802 V. (a) At finite source-drain bias, the triple points develop into triangle pairs. Relevant capacitances of the DQD can be extracted from the sizes of the hexagon and triangles. The fractional splitting, $F\sim0.2$ indicates a dominant inter-dot capacitive coupling. (b, c) Detailed bias spectroscopy of a pair of triangles at $V_{SD}$=1.0 mV and 0.5 mV. (d) Line cut along the red arrow in (b) shows resonant tunneling through excited states in the transport.}
\label{fig3}
\end{figure}

From the dimensions of the hexagon and triangles in Fig.~3(a) we
obtain the key capacitances defining the system~\cite{Wiel2002},
namely: the total capacitances of the left and right dots,
$C_{\Sigma,left(right)}$; the mutual capacitance between the two
dots, $C_m$; the relative capacitances between each side barrier
gate and its immediate neighboring dot, $C_{BL(BR),left(right)}$;
and the cross capacitance between each side barrier gate and the
next neighbouring dot, $C_{{\times}BL({\times}BR),right(left)}$.
These results agree well with modelling performed using FastCap (see
Table I). We note that by appropriate tuning of the barrier gate
voltages, we are able to form approximately symmetric dots. With the
relevant capacitances defined, we obtain the interaction energy
between the two dots, using
$E_m=(e^2/C_m)((C_{\Sigma,left}{\cdot}C_{\Sigma,right}/C_m^2)-1)^{-1}\sim$500~$\mu$eV~\cite{Mason2004}.
While the current structure enabled the formation of two nearly
identical dots, our group is developing a multi (3) layer structure,
where top-gates independently control the islands, a second layer of
gates provides contacts to in-diffused source and drain, and a third
layer provides the barrier gates. This structure allows the source
and drain reservoirs to remain populated even for low electron
numbers in the dots.

\begin{table}
\caption{Comparison of experimental values obtained from Fig. 3(a)
and modelled FastCap capacitances. For definitions,
see text.} 
\centering 
\begin{tabular}{l c c c c}
\hline\hline 
&&Experimental&  &FastCap \\ [1ex] 
&&Left(Right)&  &Modeling \\ [1ex]
\hline 
$C_{\Sigma,left(right)}$&(aF) & 22.8(26.4) &  & 30.0 \\ [1ex]
$C_{BL(BR),left(right)}$ & (aF) & 5.7(5.5) &  & 5.3 \\ [1ex]
$C_{{\times}BL({\times}BR),right(left)}$ & (aF) & 0.75(0.90)&& 0.71\\ [1ex]
$C_m$ & (aF) & 1.9 &  & 1.5 \\ [1ex]
\hline 
\end{tabular}
\label{tableI} 
\end{table}

Figures 3(b,c) show fine scans of bias triangles at $V_{SD}$=1.0 mV
and 0.5 mV respectively. Resonant tunneling through the ground state
and excited states of the double dot is clearly observed in the
high-resolution bias-spectroscopy. With increasing $V_{SD}$, the
triangular conducting regions become larger with more discrete
levels in the bias window and the overlap of the triangle pairs
increases. Figure 3(d) shows a plot of $I_{SD}$ as a function of
detuning energy, $\varepsilon$~\cite{alphafactor} between levels of
the double dot. This $I_{SD}$ line trace is extracted from a cut of
the bias triangle as shown in Fig. 3(b), where the ground and
excited state resonances are indicated by the labels \emph{a-d}. The
energy splitting of the first excited state \emph{b} to its ground
state \emph{a} is $\sim$300  $\mu$eV. We roughly estimate the
average energy-level spacing of a dot via Weyl's formula,
${\Delta}E=2\pi\hbar^2/gm^*A$, where $A$ is the area of the dot. For
a 2DEG system in Si, the effective mass of the electrons
$m^*$=0.19$m_e$ and the degeneracy $g$=4, taking into account the
spin and valley degeneracies~\cite{Magder2000}. Using this we
calculate ${\Delta}E\sim$400 $\mu$eV, which would be the expected
average level spacing if all symmetries are broken. Since no field
is applied to the dots, the spacing would be a factor of 2 larger or
$\sim$800 $\mu$eV. In Fig. (d), we monitor transport through a
serial configuration of two dots along the line cut presented in
(b). In this case, we move the energy levels in both dots in
opposite direction with respect to each other~\cite{Wiel2002} with
results in an effective reduction of a factor of two in the
experimentally expected level splitting, in good agreement with the
experimental data.

In conclusion, we have presented a tunable double-gated DQD defined
in intrinsic silicon. The fabrication of the device is reproducible
and MOS-compatible, enabling scale-up or integration into more
complex designs. Transport measurements have been performed and
extracted device capacitances were in good agreement with FastCap
modelling. High resolution bias spectroscopy of the double dot
presented evidence of resonant tunneling through ground and excited
states, indicating that the system was in the few-electron regime.
To reduce the electron number to a single electron in each dot we
propose the incorporation of additional plunger gates, independently
controlling each dot, together with an integrated charge
detector~\cite{Reilly2008} to monitor the dot occupancies. Such
Si-based double quantum dot structures would have excellent
potential for the investigation of the singlet-triplet two-level
system due the long spin-coherence times in silicon.

The authors thank D.~Barber and R.P.~Starrett for technical support
in the National Magnet Laboratory at UNSW, E.~Gauja for the
assistance in the UNSW Semiconductor Nanofabrication Facility,
C.C.~Escott, K.W.~Chan and H.~Yang for the help with the FastCap
modeling, and M.A.~Eriksson, F.A.~Zwanenburg and L.D.~Macks for helpful comments
with the manuscript. This work was supported by the Australian
Research Council, the Australian Government, and by the U.S.
National Security Agency (NSA) and U.S. Army Research Office (ARO)
(under Contract No. W911NF-04-1-0290).

\end{document}